# A Classical Approach in Simple Nuclear Fusion Reaction $_1H^2 + {_1H^3}$ using Two-Dimension Granular Molecular Dynamics Model

Sparisoma Viridi[a,*], Rizal Kurniadi[a], Abdul Waris[a], and Yudha Satya Perkasa[b]

[a]Nuclear Physics and Biophysics Research Division
Institut Teknologi Bandung, Jalan Ganesha 10, Bandung 40132, Indonesia
[b]Doctoral Program in Physics
Institut Teknologi Bandung, Jalan Ganesha 10, Bandung 40132, Indonesia
*Email: dudung@fi.itb.ac.id

Abstract

Molecular dynamics in 2-D accompanied by granular model provides an opportunity to investigate binding between nuclei particles and its properties that arises during collision in a fusion reaction. A fully classical approach is used to observe the influence of initial angle of nucleus orientation to the product yielded by the reaction. As an example, a simplest fusion reaction between $_1H^2$ and $_1H^3$ is observed. Several products of the fusion reaction have been obtained, even the unreported ones, including temporary $_2He^4$ nucleus.

## Introduction

Classical approach to investigate simple fusion reaction, where nucleus particles are considered as granular particles, is an interesting topic to conduct, as the reported classical approach in modeling He atom [1]. Granular forces scheme [2] is used instead of potential representation. This scheme accompanied by molecular dynamics method is already common in simulating small scale physical system like nanoparticles [3]. So, it is interesting to apply the same simulation scheme to smaller scale physical system: a nucleus, will fully classical approach. There are three types of force considered in this work, which are repulsive force, electrostatic force, and strong-force-like attractive force.

## Simulation

Repulsive force $\vec{R}_{ij}$, electrostatic force $\vec{Q}_{ij}$, and strong-force-like attractive force $\vec{B}_{ij}$ are considered forces in the simulation, which assumes that all nucleus particles, i.e. proton and neutron, behave like classical particles. The first force prohibits two nucleus particles to coincide, that could be similar to Pauli's exclusion principle for electrons. The second is Coulomb force acted between charged particles. And the last is binding force that hold the nucleus particles together inside the nucleus.

The repulsive force has formulation which is known as linear dash-pot model [2]

$$\vec{R}_{ij} = k_R \xi_{ij} \hat{r}_{ij}, \quad (1)$$

where $\xi_{ij}$ is defined as overlap between two nucleus particles, which is

$$\xi_{ij} = \max\left[0, \frac{1}{2}(d_i + d_j) - r_{ij}\right]. \quad (2)$$

The electrostatic force has formulation

$$\vec{Q}_{ij} = \hat{r}_{ij}\left(k_Q \frac{q_i q_j}{r_{ij}^2}\right). \quad (3)$$

And the binding force has representation in a form of

$$\vec{B}_{ij} = \hat{r}_{ij}\left(k_B \frac{b_i b_j}{r_{ij}^2}\right), \quad r_{ij} < r_B, \quad (4)$$

with $r_B$ is averaged nucleus radius, where are all the nucleus particles confined. Equations (1), (3), and (4) used following definitions

$$\vec{r}_{ij} = \vec{r}_i - \vec{r}_j, \quad (5)$$

$$r_{ij} = |\vec{r}_{ij}| = \sqrt{\vec{r}_{ij} \cdot \vec{r}_{ij}}, \quad (6)$$

$$\hat{r}_{ij} = \frac{\vec{r}_{ij}}{r_{ij}}. \quad (7)$$

In simulation the molecular dynamics method is accompanied by Gear predictor-corrector algorithm of fifth order [4], which has prediction step (written with upper index $p$) and correction step for every particular grain. The prediction step is formulated as

$$\begin{pmatrix} \vec{r}_0^{\,p}(t+\Delta t) \\ \vec{r}_1^{\,p}(t+\Delta t) \\ \vec{r}_2^{\,p}(t+\Delta t) \\ \vec{r}_3^{\,p}(t+\Delta t) \\ \vec{r}_4^{\,p}(t+\Delta t) \\ \vec{r}_5^{\,p}(t+\Delta t) \end{pmatrix} = \begin{pmatrix} 1 & 1 & 1 & 1 & 1 & 1 \\ 0 & 1 & 2 & 3 & 4 & 5 \\ 0 & 0 & 1 & 3 & 6 & 10 \\ 0 & 0 & 0 & 1 & 4 & 10 \\ 0 & 0 & 0 & 0 & 1 & 5 \\ 0 & 0 & 0 & 0 & 0 & 1 \end{pmatrix} \begin{pmatrix} \vec{r}_0(t) \\ \vec{r}_1(t) \\ \vec{r}_2(t) \\ \vec{r}_3(t) \\ \vec{r}_4(t) \\ \vec{r}_5(t) \end{pmatrix} \quad (8)$$

and the correction step as

$$\begin{pmatrix} \vec{r}_0(t+\Delta t) \\ \vec{r}_1(t+\Delta t) \\ \vec{r}_2(t+\Delta t) \\ \vec{r}_3(t+\Delta t) \\ \vec{r}_4(t+\Delta t) \\ \vec{r}_5(t+\Delta t) \end{pmatrix} = \begin{pmatrix} \vec{r}_0^{\,p}(t+\Delta t) \\ \vec{r}_1^{\,p}(t+\Delta t) \\ \vec{r}_2^{\,p}(t+\Delta t) \\ \vec{r}_3^{\,p}(t+\Delta t) \\ \vec{r}_4^{\,p}(t+\Delta t) \\ \vec{r}_5^{\,p}(t+\Delta t) \end{pmatrix} + \begin{pmatrix} c_0 \\ c_1 \\ c_2 \\ c_3 \\ c_4 \\ c_5 \end{pmatrix} \Delta \vec{r}_2(t+\Delta t), \quad (9)$$

with

$$\Delta \vec{r}_2(t+\Delta t) = \vec{r}_2(t+\Delta t) - \vec{r}_2^{\,p}(t+\Delta t). \quad (10)$$

The term $\vec{r}_n(t+\Delta t)$ is defined as

$$\vec{r}_n(t) = \frac{(\Delta t)^n}{n!}\left[\frac{d^n \vec{r}_0(t)}{dt^n}\right], \quad (11)$$

where $\vec{r}_0$ is position of a grain. The term $\vec{r}_2(t+\Delta t)$ in correction term in Equation (10) is obtained from Newton's second law of motion. For example, particle $i$ has

$$[\vec{r}_2(t+\Delta t)]_i = \frac{(\Delta t)^2}{m_i} \times \sum_{j \neq i} \vec{Q}_{ij}(t+\Delta t) + \vec{B}_{ij}(t+\Delta t) + \vec{R}_{ij}(t+\Delta t). \quad (12)$$

The left part of Equation (12) is calculated using $\vec{r}_n^{\,p}(t+\Delta t)$.

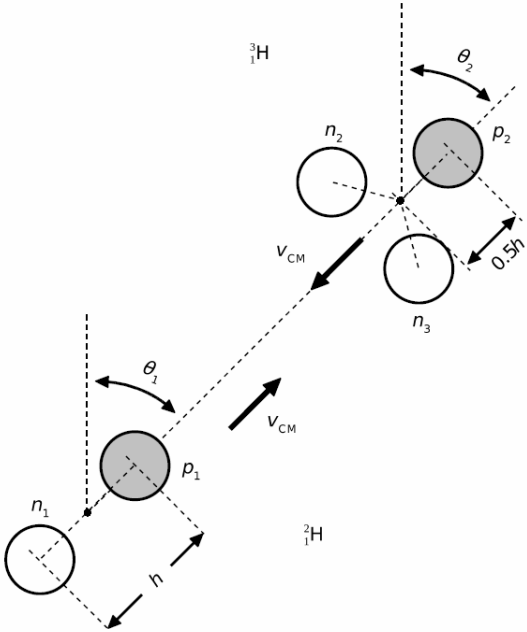

Figure 1. Initial configuration of fusion reaction between hydrogen isotopes: $_1H^2$ (lower left) and $_1H^3$ (upper right).

In order to characterize the fusion reaction which occurs through collision between hydrogen isotopes ($_1H^2$ and $_1H^3$) the problem is limited only to variation of initial orientation of the isotopes in two-dimension. Before the collision, both isotopes move in linear motion without rotation. After the collision, which depends on the initial orientation of the isotopes, some nucleus particles are interchanged.

For simplicity, the line that connects center of mass (CM) of both colliding nuclei are chosen as $x$ axis. Then the problem can be viewed as one dimensional problem, since, whether the nucleus particles bind together in a nucleus or not, can already be seen with one dimension. In Figure 1 the $n$ and $p$ indicate neutron and proton particles, respectively.

## Results and discussion

Parameters used in the simulation are: $r = 0.05$, $m = 0.1$, $q = 0, 1$, $b = 1$, $k_B = 0.2$, $r_B = 0.2$, $k_R = 10^5$, $k_Q = 0.01$, $\theta_1 = 2n\pi/16$ ($n = 0,..,15$), $\theta_2 = \pi/4$, $v = \sqrt{2}$, and $h = 0.4\sqrt{2}$. Even force representation is used in this work, it it still interesting to preset the potential of two nucleus particles. Potenstial function for the given parameters is shown in Figure 2.

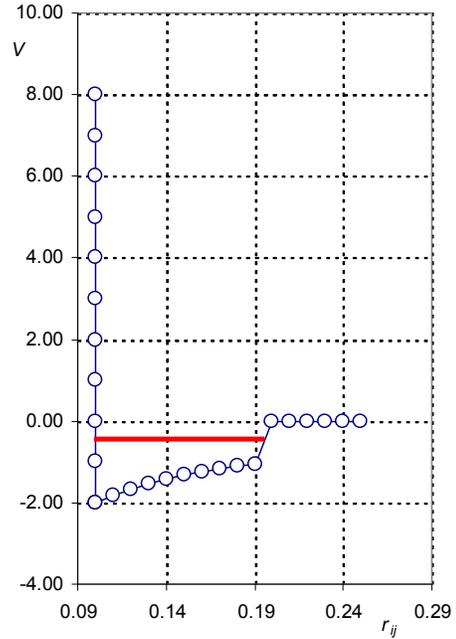

Figure 2. Potential $V$ as function of two nucleus particles separation distance $r_{ij}$, for $r = 0.05$, $m = 0.1$, $q = 0, 1$, $b = 1$, $k_B = 0.2$, $r_B = 0.2$, $k_R = 10^5$, $k_Q = 0.01$, with illustration of initial kinetic energy (drawn in red solid line).

The binding condition a two nucleus particles are determined by their potential $V$ and their kinetic energy. The potential $V$ is sum of all considered potential obtained from the forces in Equation (1), (3), and (4). Figure 2 shows the typical potential function for two nucleus particles for mentioned simulation parameters. The function $V$ is not smooth since it consists from three different forces. Intial velocity is set as the initial kinetic energy for the nucleus particles. Even the profile of $V$ is simple for a pair of nucleus particles, but it can be complecated when more that two particles are binded as occurred in the collision. It will be shown that even the orientation of a nucleus can determine difference in the product after the collision.

Typical simulation results (also the nulei before and after the collision) are shown in Figure 3 where the square mark indicates proton and the circle mark indicates neutron.

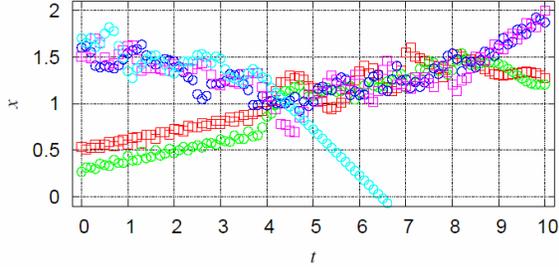

Figure 3. Position of nuclei particles in $x$ direction as function of time $t$ for $\theta_1 = \pi/4$.

As an example collision of $_1H^2$ and $_1H^3$ are shown in Figure 4. The nucleus $_1H^2$ is positioned about $x = 0.8$ and The nucleus $_1H^3$ is positioned about $x = 1.6$. Protons are drawn in square mark and neutrons are drawn in circle mark. As it can be seen during the collision there is releasing of neutron at about $t = 4.8$, while a temporary nuleus of $_2He^4$ exists about $\Delta t = 4.5$, that decay into two identical $_1H^2$ at about $t = 9$. Even with the same colliding nuclei the final results can be different when the initial orientation is different. Previous example is for $\theta_1 = \theta_2 = \pi/4$.

By varying value of $\theta_1$ several final fusion product can be obtained

$$_1H^2 + {_1H^3} \rightarrow {_1H^2} + {_1H^3}. \qquad (13.a)$$

$$_1H^2 + {_1H^3} \rightarrow {_1H^2} + {_1H^2} + {_0n^1}. \qquad (13.b)$$

$$_1H^2 + {_1H^3} \rightarrow {_1H^3} + {_1H^2}. \qquad (13.c)$$

$$_1H^2 + {_1H^3} \rightarrow {_1p^1} + {_1H^3} + {_0n^1}. \qquad (13.d)$$

$$_1H^2 + {_1H^3} \rightarrow {_1p^1} + {_1H^2} + {_0X^2}. \qquad (13.e)$$

Equation (13.a) shows the unaffected colliding nuclei, they are just colliding and remain the same after collision, while Equation (13.c), even with the same product, the nuclei interchanges their nucleus particle. Equation (13.b) produce two $_1H^2$ with one neutron as given in Figure 2. Equation (13.d) shows, whether there is nucleus particle interchange, a separation of $_1H^2$ into one proton and one neutron because hit by $_1H^3$. And the final reaction in Equation (13.e) gives unphysical yield that produce on proton, one $_1H^2$, and one $_0X^2$, which has not yet been reported. The usual results of fusion product of $_1H^2$ and $_1H^3$ is [5]

$$_1H^2 + {_1H^3} \rightarrow {_2He^3} + {_0n^1}. \qquad (14)$$

Or that $_2He^3$ is usually produced by pairs of $_1H^2$ in cold fusion [6-7].

In our simulation actually we have also found $_2He^4$ instead of $_2He^3$ but it is not stable that decays into two $_1H^2$ after first releasing a neutron.

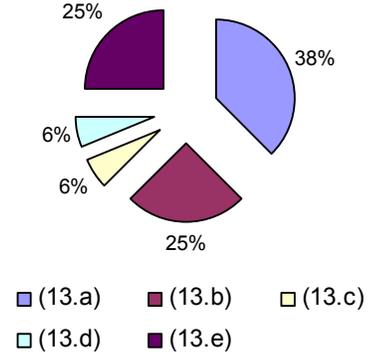

Figure 4. Probability of Equation (13.a)-(13.c) to occur for 16 variations of $\theta_1$.

Value of $\theta_1$ is varied using relation $2n\pi/16$ with $n = 0, .., 15$, there will be 16 fusion products that are already categorized into five types as in Equation (13.a)-(13.e). Each type has probability of 6/16, 4/16, 1/16, 1/6, and 4/16, respectively (shown in pie chart in Figure 4).

**Conclusion**

A classical approach in simulating a simple fusion reaction has been conducted and several fusion products, which is dependent on initial orientation of the nuclei, has been found. Unfortunately, none the products are real except the temporary $_2He^4$ nucleus. The unphysical results could be addressed to the parameters that are not set or chosen accordingly to fit the physical results.

**Acknowledgements**

Authors would like to thanks ITB Research Division research grant in 2010 and 2011 for supporting this work and presentation in the conference.